\documentclass[twocolumn,aip,amsmath,amssymb,reprint,citeautoscript]{revtex4-1}  
\usepackage{graphicx}  
\usepackage{dcolumn}   
\usepackage{bm}        
\usepackage{amssymb}   
\usepackage{hyperref}
\usepackage{amsmath}
\usepackage{mathtools}
\usepackage{float}
\usepackage{color}
\usepackage{soul}

\allowdisplaybreaks

\newcommand{\citeasnoun}[1]{Ref.~\citenum{#1}}
\newcommand{\citeasnounp}[1]{Refs.~\citenum{#1}}
\renewcommand{\vec}[1]{\mathbf{#1}}
\newcommand{\figref}[1]{Fig.~\ref{#1}}

\newcommand{\equref}[1]{Eq.~(\ref{#1})}
\newcommand{\secref}[1]{Sec.~\ref{#1}}

\begin{document}

\title{Nonlinear exceptional-point lasing with ab-initio Maxwell-Bloch theory}

\author{Mohammed Benzaouia}
\thanks{Corresponding author: medbenz@stanford.edu\\ Currently at Ginzton Lab, Stanford.}\affiliation{Department of Electrical Engineering and Computer Science, Massachusetts Institute of Technology, Cambridge, MA 02139, USA.} 
\author{A. D. Stone} \affiliation{Department of Applied Physics, Yale University, New Haven, CT 06520, USA.}
\author{Steven G. Johnson} \affiliation{Department of Mathematics, Massachusetts Institute of Technology, Cambridge, MA 02139, USA.}

\begin{abstract}
We present a general analysis for finding and characterizing nonlinear exceptional point (EP) lasers above threshold, using steady-state \emph{ab-initio} Maxwell-Bloch equations. For a system of coupled slabs, we show that a nonlinear EP is obtained for a given ratio between the external pumps in each resonator, and that it is associated with a kink in the output power and lasing frequency, confirming coupled-mode theory predictions. Through numerical linear stability analysis, we confirm that the EP laser can be stable for a large enough inversion relaxation rate. We further show that the EP laser can be characterized by scattering a weak signal off the lasing cavity, so that the scattering frequency spectrum exhibits a \emph{quartic} divergence. Our approach can be applied to arbitrary scatterers with multi-level gain media.
\end{abstract}
\maketitle

\section{Introduction}

Exceptional points (EPs) are special singularities of certain non-Hermitian systems where at least two eigenvalues and their eigenvectors coalesce and become degenerate~\cite{heiss2012physics, miri2019exceptional}. EPs have attracted much attention recently in photonics~\cite{miri2019exceptional}, where gain and loss can be used to tailor the non-hermiticity. At these singularities the relevant system operator is not diagonalizable anymore, leading to a response typically quite different from that of Hermitian systems. EPs have been mainly studied in the context of linear photonic systems. In laser theory, the lasing frequencies are the solution of a nonlinear eigenvalue problem~\cite{esterhazy2014scalable} (except at threshold) and there has been significant interest in studying the analog of linear EPs for this physically important type of nonlinear systems. In this Letter, we perform a comprehensive analysis for finding and characterizing EP lasers above threshold.

The influence of EPs on lasers has already been explored with respect to certain lasing properties, such as demonstrating lasing turn-off with increasing pump power~\cite{liertzer2012pump, brandstetter2014reversing, peng2014loss}, the enforcement of single-longitudinal mode behaviour~\cite{hodaei2014parity, feng2014single}, and realization of unidirectional chiral lasers~\cite{peng2016chiral, longhi2017unidirectional, miao2016orbital}. \citeasnoun{kim2016direct} also presented experimental evidence of an EP above threshold in coupled photonic-crystal lasers. However, nonlinear effects for a laser that is \emph{exactly} at the EP above threshold were only included in few previous works with certain approximations (e.g., zero-dimensional rate equations for coupled lasers~\cite{kominis2017spectral, kominis2018exceptional}) or limited physical interpretation~\cite{liertzer2015exceptional}. In this work, we perform a comprehensive study of EP lasers above threshold using \emph{ab-initio} Maxwell--Bloch equations, complemented with a simpler nonlinear CMT model. We also propose and analyze a way to experimentally observe the EP by operating the laser as a parametric amplifier, in which case we predict enhanced sensitivity at small frequency shifts, with potential use as a novel sensor. 

For a system with two coupled resonators, as we show by CMT (\secref{sec:CMT}) and confirm by steady-state Maxwell--Bloch solution (\secref{sec:SALT}), a nonlinear lasing EP occurs at a specific ratio $D_2/D_1$ of the pump in each resonator, which depends on the loss rate and coupling strength. For such a laser, the EP corresponds to a transition between a PT-broken phase (where the lasing mode resides mainly on a single cavity) and a PT-symmetric phase (where the mode is spread in both cavities). We also show that this EP laser can be stable for a large enough population-inversion relaxation rate, and is characterized by a kink (discontinuity of derivative) in the laser frequency and output power. Our analysis further shows that the previously observed laser turn-off with increasing pump~\cite{liertzer2012pump, brandstetter2014reversing, peng2014loss} can be directly attributed to a \emph{virtual} nonlinear EP. Finally, as noted above, we demonstrate that the EP laser behaviour can also be characterized by a scattering experiment, where the transmission frequency-spectrum of a weak signal field scattered off the laser cavity exhibits a \emph{quartic} divergence (\secref{sec:scattering}). 

\section{Coupled mode theory (CMT)}
\label{sec:CMT}

Although we look at a more a complete model later, we start here by studying a simple dimer laser using a CMT model with two scalar degrees of freedom. For two coupled resonators with nonlinear gain saturation, steady state solutions can be described by~\cite{hassan2015nonlinear, ge2016nonlinear, zhu2019laser} 
\begin{equation}\label{eq:cmt}
    \begin{pmatrix}i(G_1-\kappa_1) && g \\ g && i(G_2-\kappa_2)\end{pmatrix} \begin{pmatrix}\psi_1 \\ \psi_2 \end{pmatrix} = \Delta\begin{pmatrix}\psi_1 \\ \psi_2 \end{pmatrix},
\end{equation}

We show that the lasing solutions (real $\Delta$) for this coupled system of nonlinear equations are of two types (see details in the Supplementary Material; similar results to ours can be recovered from \citeasnounp{hassan2015nonlinear, ge2016nonlinear} in special cases). First, there are PT-symmetric modes, characterized by
\begin{equation}\label{eq:modes1}
    \Delta = \pm \sqrt{g^2-\left|\frac{\kappa_2D_1-\kappa_1D_2}{D_1+D_2}\right|^2}, \;\; G_1-\kappa_1 = -(G_2-\kappa_2).
\end{equation}
One sees that these modes are detuned from the resonator frequency, with either sign of detuning possible. The second equation above implies that the resulting operator has PT symmetry: the net gain in one cavity equals the net loss in the other so that interchanging the cavities and complex conjugating leaves the operator unchanged. The squared modal amplitude is the same in each cavity  and given by $|\psi_n|^2=(D_1+D_2)/(\kappa_1+\kappa_2)-1$. The pumps at threshold  $\left(|\psi_n|=0\right)$ do not need to be equal, but must satisfy $D_1^{th}+D_2^{th}=\kappa_1+\kappa_2$. 

There is also a PT-broken mode characterized by
\begin{equation}\label{eq:modes2}
    \Delta = 0, \;\; (G_1-\kappa_1)(G_2-\kappa_2)+g^2=0,
\end{equation}
with zero detuning, and with the corresponding condition at threshold $(D_1^{th}-\kappa_1)(D_2^{th}-\kappa_2)+g^2=0$. This implies that one cavity has net gain and the other net loss, but they need not be balanced, and the lasing mode will reside mainly in the gain cavity, with only small penetration into the lossy cavity. The limit $g\to 0$ corresponds to independent lasing in each cavity. 
Together these two solutions give the following picture (Fig. 1): If the pump is increased in one of the cavities maintaining a large enough difference between the net gain in the two cavities, the laser turns on in a single, highly asymmetric mode, with a corresponding asymmetric emission pattern; if the pumps are increased fairly equally from below threshold, the laser will turn on a single mode with equal intensity in each cavity, which is locked to the PT-symmetry condition, and which will have more symmetric emission pattern (equal if the radiative loss rates are equal).

\begin{figure}
    \includegraphics[width=1\columnwidth, keepaspectratio]{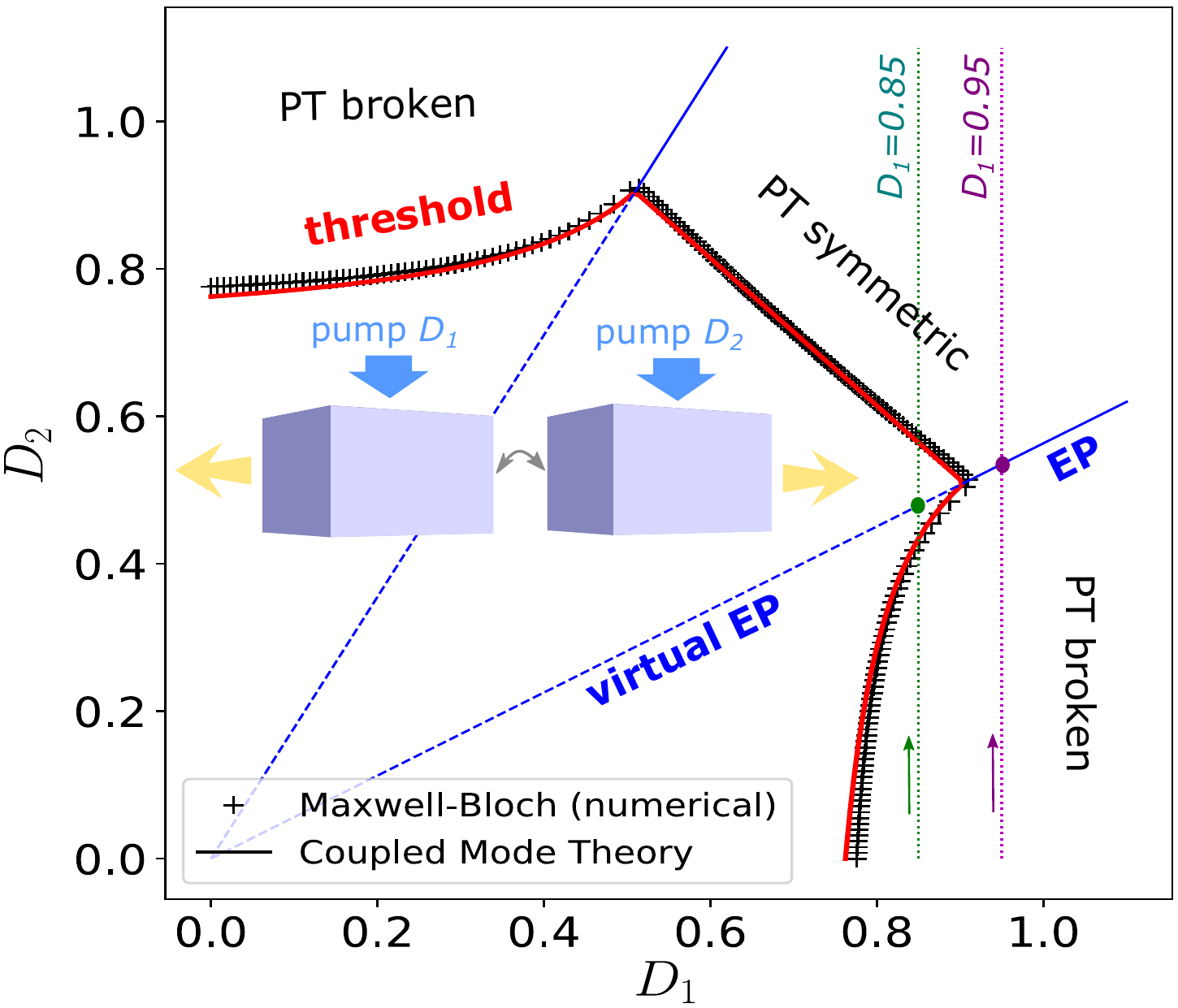}
    \caption{Phase diagram for a laser consisting of two coupled slabs with independent external pumps $D_1$ and $D_2$. The two-level gain medium inside the slabs is characterized by a central frequency $\omega_aa/2\pi c=1$ and linewidth $\gamma_\perp a/2\pi c=0.1$. Each slab has a refractive index $n_c=3$, conductivity loss $\sigma_c=0.5\omega_a$ and thickness $t=5a/n_c$, with an air gap $d\approx0.2174a$, where $a$ is an arbitrary unit length. `+' marker (resp. red solid line) shows the numerical Maxwell--Bloch equation (resp. CMT) threshold. A good fit is obtained using a loss factor $\kappa \approx 0.71$ and coupling coefficient $g \approx 0.2$. Blue solid (dashed) lines show the EP (virtual EP) as predicted by \equref{eq:EP_condition}. Magenta (green) dotted line shows the pump path passing through the EP [\figref{Fig_power}] (virtual EP [Supplementary Material]).}
     \label{Fig_threshold-map}
\end{figure}

\smallskip
The transition between the two phases occurs at
\begin{equation}\label{eq:EP_condition}
    D_2 = \frac{\kappa_2\mp g}{\kappa_1\pm g}D_1,
\end{equation}
which requires $\kappa_{1,2}\geq g$. When crossing each of these boundaries, the two PT-symmetric modes coalesce at a \emph{single} solution, then transition into the PT-broken phase with a lasing mode and an additional non-lasing solution (with complex $\Delta$). The transition then corresponds to a nonlinear EP, with
\begin{equation}
    \Delta_{\text{EP}}=0, \; \; \psi_{\text{EP}} = e^{i\theta} \sqrt{\frac{D_1}{\kappa_1\pm g}-1}\; \begin{pmatrix} 1 \\ \mp i \end{pmatrix} (\theta\; \text{arbitrary}).
\end{equation}
The $\pm \pi/2$ phase difference between the amplitudes $\psi_n$ in each resonator indicates a chiral behaviour~\cite{heiss2001chirality}. It is important to note that these two EPs are actual solutions $\left(|\psi_\text{EP}|>0\right)$ only when $D_1>\kappa_1\pm g$. One way to obtain the EP would then be to first increase the pump $D_1$ above the required value, then to increase the pump $D_2$ up to the value set by \equref{eq:EP_condition}. We can also keep tracking the nonlinear EP along the line defined by \equref{eq:EP_condition}, by simultaneously changing pumps $D_1$ and $D_2$. 

When the previous condition on $D_1$ is not satisfied, we can still have a \emph{virtual} EP, which corresponds to a formal solution of \equref{eq:cmt} but with \emph{negative} intensity [i.e., $G_n=D_n/(1-|\psi_n|^2)$]. Of course, this does not correspond to a physical solution since the laser turns off when $|\psi_n|=0$. However, formally tracking this mathematical solution allows for a very nice interpretation of the suppression and revival of lasing~\cite{liertzer2012pump, brandstetter2014reversing, peng2014loss}, which can now be understood as a \emph{virtual} nonlinear EP, as we show later. 

We can also compute the output power as $P_{out} = \kappa_{1,r}|\psi_1|^2 + \kappa_{2,r}|\psi_2|^2$, where $\kappa_{n,r}$ denotes the radiative loss rate.  In particular, we find that, when transitioning between the PT-symmetric and PT-broken phases, there is a kink (discontinuous first derivative) in $P_{out} (D_2)$, due to the switch betwen the two types of solutions.  For simplicity, we compute this derivative at the EP assuming equal loss rates $\kappa_1=\kappa_2 \equiv \kappa$ in the two cavities (and maintain this assumption from here on),
\begin{equation}
    \frac{\partial P^{\textrm{brok}}_{out}}{\partial D_2} = \frac{\kappa_r\left[\kappa D_1-(\kappa\pm g)^2\right]} {\kappa^2D_1-(\kappa^2-g^2)(\kappa\pm g)}, \;\; \frac{\partial P^{\textrm{sym}}_{out}}{\partial D_2} = \frac{\kappa_r}{\kappa}.
\end{equation}
The largest discontinuity is observed for the smallest possible $D_1$ $(=\kappa\pm g)$, in which case we have a slope $\mp\kappa_r/g$ in the PT-broken side, compared to $\kappa_r/\kappa$ in the PT-symmetric phase. 
Physically as one starts equalizing the pumps from the broken phase, the laser becomes less efficient because it has more intensity in the lossy cavity, and the power slope is negative; before the pumps become equal the laser jumps to the symmetric phase, where pumping harder now helps, and the power slope becomes positive (Fig. 2a).  In the experiments~\cite{brandstetter2014reversing, peng2014loss} where the laser turns on and off with increasing pumping, the intensity in the broken phase is driven to zero before the EP is reached (this is a virtual EP, as mentioned above); and only at higher pumps does it turn on in the symmetric phase. If the phase boundary is crossed at higher values, the discontinuity is weaker and eventually disappears in the limit $D_1\rightarrow\infty$, as both slopes are equal to $\kappa_r/\kappa$.

\section{Maxwell--Bloch equations}
\label{sec:SALT}

The previous coupled mode theory is a nice tool to obtain qualitative and semi-quantitative results, but it does not take into account other effects such as spatial mode profile and spatially varying gain saturation, gain-medium lineshape, and, importantly, gain medium dynamics which can affect the stability of the system. Here, we obtain numerical ab-initio results, within the framework of the semi-classical Maxwell--Bloch equations (with the rotating-wave approximation)~\cite{haken1986laser} that describe the interaction between the electromagnetic field and the gain medium modeled as a two-level system (for simplicity, written here assuming a single polarization)  
\begin{align*}
\nabla ^2 E^+ &=  \ddot P^+ + \epsilon_c\ddot E^+ +\sigma_c \dot E^+ \\
i\dot P^+ &= (\omega_a-i\gamma_\perp)P^+ + \gamma_\perp E^+D \stepcounter{equation}\tag{\theequation}\label{eq:MB} \\
\dot D/\gamma_\parallel &= D_0-D+\text{Im} (E^{+*}\cdot P^+),
\end{align*}
where $E^+$ is the positive-frequency component of the electric field (the physical field being given by $2\text{Re}[E^+]$), $P^+$ is the positive-frequency polarization describing the transition between two energy levels (with frequency $\omega_a$ and linewidth $\gamma_\perp$), $D$ is the population inversion (with relaxation rate $\gamma_\parallel$), $D_0$ is the pump strength profile, $\epsilon_c$ is the cold-cavity real permittivity, and $\sigma_c$ is a cold-cavity conductivity loss. Here, we are assuming that the orientation of the atomic transition is parallel to the electric field, and have written all three fields in their natural units~\cite{burkhardt2015steady}. Multi-level transitions can also be included~\cite{cerjan2012steady}. A steady-state solution of these equations can be obtained via steady-state ab-initio laser theory (SALT), which is exact (within the rotating wave approximation) for single-mode lasing and approximate for multi-mode lasing with well-separated modes~\cite{tureci2006self, ge2008quantitative, ge2011, esterhazy2014scalable}. We use it to track single-mode lasing solutions as the external pump is changed by numerically solving the nonlinear steady-state equation~\cite{esterhazy2014scalable}
\begin{equation} \label{eq:SALT}
    \Theta E_\ell \equiv \left(\nabla^2+\omega_\ell^2\left[\epsilon_c + i\frac{\sigma_c}{\omega_\ell} +\Gamma_\ell D_\ell \right]\right) E_\ell=0
\end{equation}
where $\Gamma_\ell = \Gamma(\omega_\ell) \equiv \gamma_\perp/\left(\omega_\ell-\omega_a+i\gamma_\perp \right)$ and $D_\ell = D_0 / \left(1 + |\Gamma_\ell E_\ell|^2\right)$, with a corresponding polarization $P_\ell = \Gamma_\ell D_\ell E_\ell$.

\begin{figure}
    \includegraphics[width=\columnwidth, keepaspectratio]{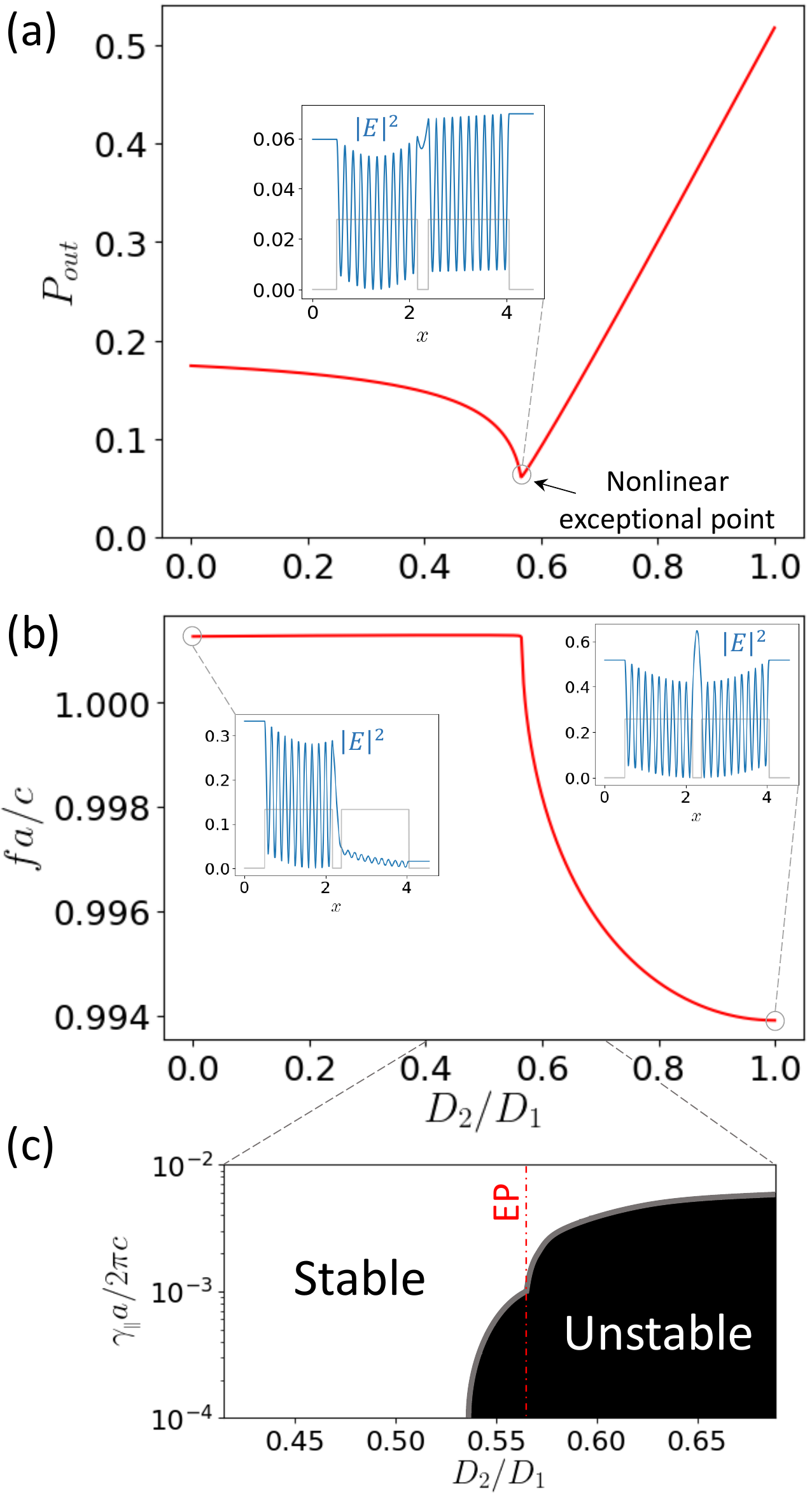}
    \caption{Output power (a) and frequency (b) for a single-mode steady state solution of the nonlinear Maxwell--Bloch equation [\equref{eq:SALT}]. We pump the first slab until $D_1=0.95$ (not shown in the plot), and then increase $D_2$ (keeping $D_1$ fixed). Insets show the mode intensity profile for different values of $D_2$. (c) shows results of the numerical stability analysis. The nonlinear EP laser can be stable for large enough $\gamma_\parallel$.}
    \label{Fig_power}
\end{figure}

It is important to realize that the operator $\Theta$ depends nonlinearly on spatially varying $E_\ell$ due to gain saturation and hole burning above threshold. In particular, even if the system has an EP in the linear regime below threshold ($E_\ell \rightarrow 0$), the EP will typically disappear above threshold since the operator changes. So, the parameters of the system need to be carefully chosen and varied to take into account these nonlinearities and obtain a lasing EP above threshold.

\smallskip
We now consider a 1d system formed by two coupled slabs with non-saturated pump strengths $D_1$ and $D_2$ as a realization of the system described by CMT (see inset and caption of \figref{Fig_threshold-map}). We first find the threshold map giving the required pumps $(D_1^{th},D_2^{th})$ for lasing. Results are shown in \figref{Fig_threshold-map}. The map, where the two different threshold curves correspond to the PT-symmetric and PT-broken phase, matches qualitatively previous experimental results~\cite{brandstetter2014reversing} and coupled mode theory analysis~\cite{zhu2019laser}. Here, we also confirm that the full-wave numerical solutions can be accurately described by CMT.
A good fit here is obtained by taking the CMT parameters $\kappa_1=\kappa_2=\kappa \approx 0.71$ and $g \approx 0.2$. 

Following the CMT analysis, we can now try to observe a nonlinear EP laser by first increasing $D_1$ to a fixed value, then increasing $D_2$ to the value given by \equref{eq:EP_condition}. It is important to note that the CMT results are semi-quantitative. So, in order to obtain a good EP from the Maxwell--Bloch equations, simply changing the pump for an arbitrary structure won't reach an EP. Instead we need to fine-tune one structural parameter; here, we fine-tune the spacing $d$ between the two slabs and focus on the first EP. 

We pump the first slab until $D_1=0.95$, which, as required, is larger than $\kappa+g\approx 0.91$. We then keep $D_1$ fixed and increase the pump $D_2$ along the line shown in Fig.1. Results obtained from numerically solving the nonlinear \equref{eq:SALT} [using a finite-difference scheme] are shown in \figref{Fig_power}(a,b). We see that, at first, the output power steadily decreases for increasing $D_2$ with an approximately constant laser frequency $f$. This corresponds to the PT-broken phase discussed in the CMT section. This is clearly confirmed by the asymmetric mode profile shown in the inset of Fig. 2b for $D_2=0$. The system then reaches an EP, clearly characterized by a kink in the output power and the laser frequency. The EP occurs for $D_2/D_1\approx0.566$, which matches the expectation from CMT given by $(\kappa-g)/(\kappa+g)\approx0.56$ with the parameters derived from fitting the threshold map. At higher pump values, the output power increases almost linearly with the pump, as predicted by the CMT modal amplitude in the PT-symmetric phase. While we only show one lasing mode, a second lasing mode can also be found in the PT-symmetric region (see Supplementary Material). Experimental evidence of a laser-frequency kink was indeed observed in \citeasnoun{kim2016direct} using coupled photonic-crystal cavities, and our results can be used to interpret the experiments (see Supplementary Material). There, tuning $D_2/D_1$ was achieved by changing the position of the optical pump. A graphene sheet was then used to change the loss of one of the cavities and tune the EP position as given in \equref{eq:EP_condition}. A similar output power kink was also experimentally observed in \citeasnoun{brandstetter2014reversing}. As discussed above, while the EP (and frequency-kink) is present even at higher pumps, the output power kink is most manifest when the EP is near threshold. 
\smallskip

As noted above, during the first stage of increasing $D_2$, the output power steadily decreases to reach a minimum value at the EP. This minimal value becomes smaller as $D_1$ is decreased, and reaches zero when $D_1=\kappa+g$. When $D_1$ takes smaller values, the laser shuts off before reaching the nonlinear EP. However, formal mathematical solutions of \equref{eq:SALT} with negative effective power [$D_\ell = D_0 / \left(1 - |\Gamma_\ell E_\ell|^2\right)$] can still be tracked~\cite{cerjan2019multimode}, and a corresponding \emph{virtual} EP can still be found (see green dotted line in Fig. 1 and additional plots in Supplementary material for behaviour with $D_1=0.85<\kappa+g$). This gives additional insight into the suppression and revival of lasing~\cite{liertzer2012pump, brandstetter2014reversing, peng2014loss}, as due to the presence of a \emph{virtual} nonlinear EP.  Specifically, a negative power slope of the laser with increasing pump, as shown in Fig. 2c, will continue to zero intensity, turning the laser off; it will then turn on at even higher pump without passing through an EP, and there will be no measured kink in the power slope.
\smallskip

 Finally, a crucial point that should be studied carefully is the stability of the lasing solution at the EP. Rigorous stability analysis has been previously done by linearizing the Maxwell--Bloch \equref{eq:MB} around a single-mode lasing solution~\cite{burkhardt2015steady, liu2017symmetry, benzaouia2020single}, which leads to a quadratic eigenvalue problem that can be solved numerically (or semi-analytically close to threshold). We perform a similar numerical stability analysis on the EP lasing solution, and show the results in \figref{Fig_power}(c) as a function of $\gamma_\parallel$. Importantly, we see that the EP solution is stable for a large enough value of $\gamma_\parallel$ (here, $\gamma_\parallel a/2\pi c\gtrsim 10^{-3}$). Similar stability plots were also found in \citeasnoun{liertzer2015exceptional}. The linearized Maxwell-Bloch equations (stability eigenproblem) also confirm the presence of an EP as a coalescence of two eigensolutions (see Supplementary Material). 

\section{Parametric amplifier}
\label{sec:scattering}

 We have seen that the nonlinear EP is typically associated with a kink in the lasing frequency and output power (at least for a small enough $D_1$). However, even if the system is only \emph{near} the EP, the previous trends are expected to persist, but the kinks are changed into smoother variations. We show here that we can characterize and confirm the EP using a scattering setup. In particular, we scatter light off the laser cavity using an infinitesimal external source at frequency $\omega_s$ with a detuning $\sigma=\omega_s-\omega_\ell$. We can compute the scattered field by linearizing the Maxwell--Bloch \equref{eq:MB} around the lasing solution. In this limit, a degenerative four-wave mixing process occurs leading to an idler field at frequency $\omega_i = 2\omega_\ell-\omega_s$. The fields can then be obtained by numerically solving the following coupled equations, where subscripts $s$ and $i$ refer to signal and idler fields (see details in Supplementary Material)
 
 \begin{widetext}
 \begin{equation}\begin{split}
    -i\omega_s\;\delta J_s&=\nabla^2 \delta E_s +\omega_s^2\left[\epsilon_c+\frac{i\sigma_c}{\omega_s}+\Gamma_s D_\ell \left(1+\alpha |E_\ell|^2\right)\right]\delta E_s+\omega_s^2 \Gamma_s D_\ell \beta E_\ell^2\; \delta E_i^*,\; \alpha = \frac{\Gamma_s-\Gamma_\ell^*}{2\left(i+\sigma/\gamma_\parallel\right)+\left(\Gamma_i^*-\Gamma_s\right)|E_\ell|^2},\\
    0&=\nabla^2 \delta E_i +\omega_i^2\left[\epsilon_c+\frac{i\sigma_c}{\omega_i}+\Gamma_i D_\ell \left(1+\beta^* |E_\ell|^2\right)\right]\delta E_i+\omega_i^2 \Gamma_iD_\ell \alpha^* E_\ell^2\; \delta E_s^*,\; \beta = \frac{\Gamma_\ell-\Gamma_i^*}{2\left(i+\sigma/\gamma_\parallel\right)+\left(\Gamma_i^*-\Gamma_s\right)|E_\ell|^2}.
\end{split}\label{eq:MB-scattering} \end{equation}
\end{widetext}

We first note that in the limit $\omega_s \rightarrow \omega_\ell$, we have $\omega_i \rightarrow \omega_\ell$ and $\alpha \rightarrow \beta$. In this limit, $(\delta E_s,\delta E_i) \rightarrow (E_\ell,-E_\ell)$ is a source-free ($\delta J_s\rightarrow 0$) solution of \equref{eq:MB-scattering}, which means that $\omega_s \rightarrow \omega_\ell$ is a pole of the associated Green's function. So, this setup can be used to probe the presence of an EP by studying the slope rate of the scattered signal as a function of $\omega_s$. In reality, the divergent response would be cut off by saturation of amplification of the scattered field, but here we suppose that we are far from this regime by assuming weak signal and idler fields, and sufficient detuning $\sigma$. 

\begin{figure}
    \includegraphics[width=\columnwidth, keepaspectratio]{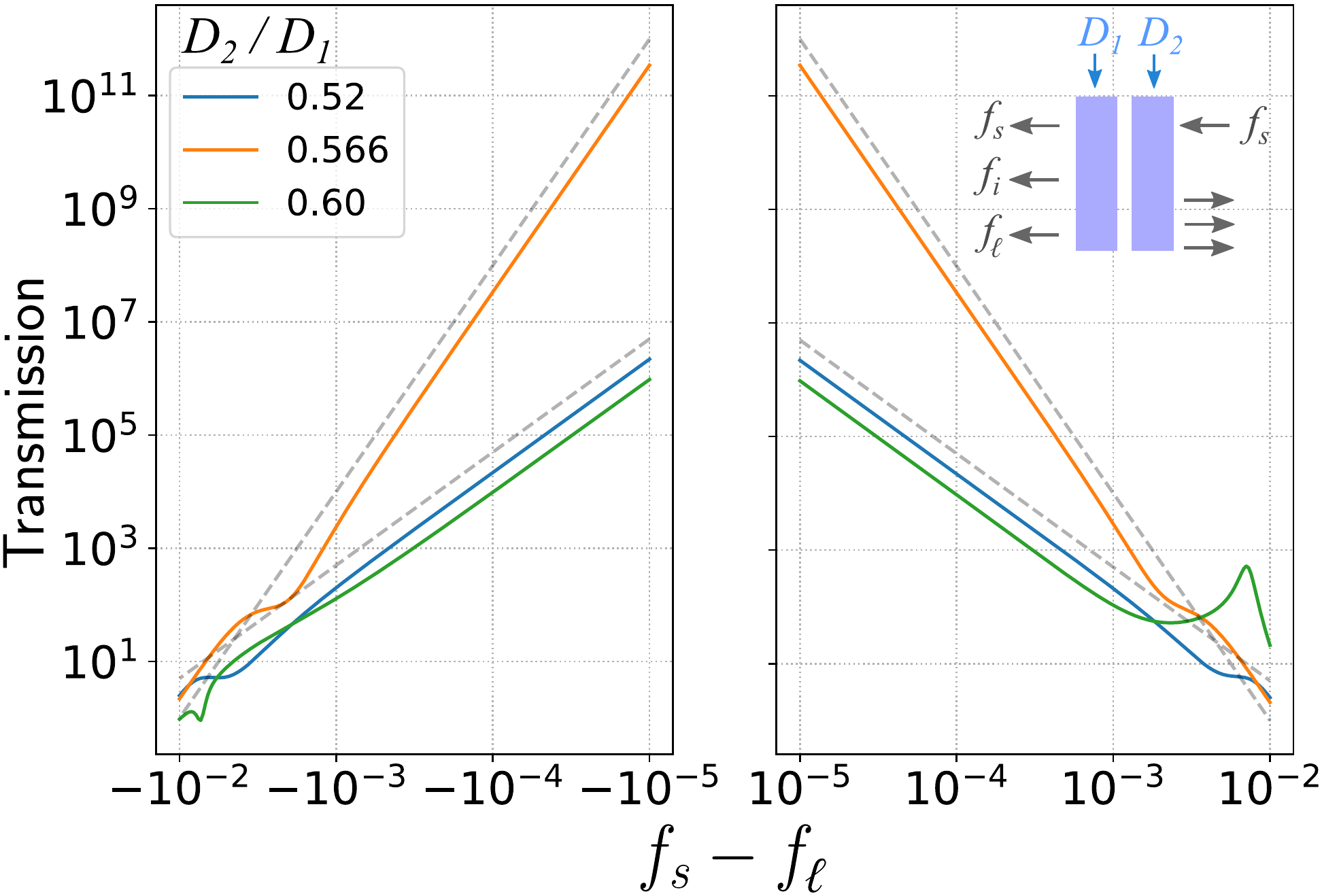}
    \caption{Transmission at the signal frequency $f_s$ due to an infinitesimal source for different pump values $D_2$ ($D_1$ being fixed at 0.95) and $\gamma_\parallel a/2\pi c=10^{-2}$. Results are obtained by numerically solving \equref{eq:MB-scattering} using a finite difference scheme. Grey dashed line represents quadratic and quartic scaling rates. The EP laser ($D_2/D_1\approx 0.566$) is clearly associated with an inverse quartic variation of the transmitted intensity with detuning from the laser line.}
    \label{Fig_scattering}
\end{figure} 

In the limit $\gamma_\parallel \rightarrow 0$, we have $\alpha,\beta \rightarrow 0$, so signal and idler fields become decoupled, and \equref{eq:MB-scattering} becomes identical to the steady-state Maxwell--Bloch \equref{eq:SALT}. This means that the associated spectral response will exhibit an EP behaviour characterized by a \emph{quartic} divergence, instead of the usual quadratic behaviour associated with a single resonance. Working at small but finite $\gamma_\parallel$ only leads to a small change in the operator of \equref{eq:MB-scattering}, so the structural parameters can be easily fine-tuned to exhibit an EP response at a given non-zero $\gamma_\parallel$. The structure presented here was indeed already chosen for a laser-scattering EP at $\gamma_\parallel a/2\pi c = 10^{-2}$; a value at which the EP laser is stable [\figref{Fig_power}(c)]. To confirm this behaviour, we numerically solve the coupled equations [\equref{eq:MB-scattering}] using a signal source on the right of the laser at frequency $f_s=\omega_s/2\pi$, and show the transmitted signal for different pump values (\figref{Fig_scattering}). We first note a diverging transmission at $f_s\rightarrow f_\ell$ associated with the (lasing) pole of the system. More importantly, we clearly see that the EP laser ($D_2/D_1\approx 0.566$) is associated with a \emph{quartic} rate, compared to a quadratic rate for the non-EP laser (e.g., $D_2/D_1\approx 0.52, 0.61$ shown in the figure). Such parametrically stronger frequency variation makes EP amplifiers potentially interesting for improved sensors, but a quantitative analysis requires a full noise theory of such amplifiers and lasers. 

\section{Conclusion and outlook}

The results presented here describe general properties of stable nonlinear EP lasers above threshold, and represent general guidelines for experimental demonstration. It can also be relevant beyond optics, such as in nonlinear PT-symmetric circuits recently considered for wireless power transfer~\cite{assawaworrarit2017robust}. One remaining theoretical challenge for future work is the computation of the EP laser linewidth by introducing noise in the model. While standard formulae~\cite{chong2012general, pick2015ab} diverge, a proper treatment should yield finite values~\cite{pick2017general} and nonlinear saturation effects will certainly come into play in some form. More generally, while we looked at the example of coupled slabs, our \emph{ab-initio} method can be used to study and design stable EP lasers in arbitrary structures and extend previous works of linear photonics to lasers above threshold~\cite{lin2016enhanced}.

\section*{Acknowledgments}
This work was supported in part by the Simons Foundation and by the U.S. Army Research Office through the Institute for Soldier Nanotechnologies under Award No. W911NF-18-2-0048. Authors would like to thank Adi Pick for useful discussions.



\section*{References}
\bibliography{biblio}

\pagebreak
\widetext
\newpage
\begin{center}
\textbf{\Large Supplementary Material\\ Nonlinear exceptional-point lasing with ab-initio Maxwell-Bloch theory}
\end{center}
\setcounter{equation}{0}
\setcounter{figure}{0}
\setcounter{table}{0}
\setcounter{page}{1}
\setcounter{section}{0}
\makeatletter
\renewcommand{\theequation}{S\arabic{equation}}
\renewcommand{\thefigure}{S\arabic{figure}}
\section{Coupled mode theory}

We study simple coupled mode theory equations for a system of two coupled resonators with nonlinear gain saturation. Steady state solutions are described by
\begin{equation}
    \begin{pmatrix}i(G_1-\kappa_1) && g \\ g && i(G_2-\kappa_2)\end{pmatrix} \begin{pmatrix}\psi_1 \\ \psi_2 \end{pmatrix} = \Delta\begin{pmatrix}\psi_1 \\ \psi_2 \end{pmatrix},
\end{equation}
where $\Delta$ is the relative frequency, $g$ is a coupling rate, $\kappa$ is a loss rate, $G_n=D_n/(1+|\psi_n|^2)$ is a gain rate that depends on the external pump $D_n$ and the field amplitude $\psi_n$.

We define $X_n=G_n-\kappa_n$. Solutions of the problem satisfy
\begin{equation}
2\Delta = i(X_1+X_2)\pm\sqrt{4g^2-(X_1-X_2)^2}, \;\;\; |\psi_n|^2=\frac{D_n}{\kappa_n+X_n}-1, \;\;\; \psi_n(iX_n-\Delta)+g\psi_m=0\; (n\neq m).
\end{equation}
Lasing solutions require a real $\Delta$. Two types of solutions are possible. First, for $X_1+X_2=0$ and $|X_1|\leq g$, we have real $\Delta = \pm\sqrt{g^2-X_1^2}$. At threshold, we have $\psi_n=0$, so $D_1^{th}+D_2^{th}=\kappa_1+\kappa_2$. We call these PT-symmetric solutions since loss and gain are balanced between the two resonators ($X_1=-X_2$). The second type is for $\Delta = 0$, which requires $X_1X_2+g^2=0$. (We call these PT-broken solutions since the gain/loss balance is lost.) When such condition holds, there is another purely imaginary solution $\Delta = i\left(X_1-g^2/X_1\right)$. The stability of the system ($\text{Im}\Delta\leq 0$) requires $X_1 \leq g^2/X_1$ ($\Leftrightarrow X_1\leq-g$ or $0\leq X_1\leq g$). At threshold, these modes satisfy $(D_1^{th}-\kappa_1)(D_2^{th}-\kappa_2)+g^2=0$. 

The previous solutions are directly determined by the value of $X_1$. For each type of solutions, $X_1$ can be found using
\begin{equation}\label{eq:x1dependence}
    \frac{D_2}{\kappa_2+X_2}-1=|\psi_2|^2=\left|\frac{\Delta-iX_1}{g}\psi_1\right|^2=\frac{X_1^2+\Delta^2}{g^2}\left(\frac{D_1}{\kappa_1+X_1}-1\right).
\end{equation}

For PT-symmetric solutions, \equref{eq:x1dependence} simplifies to
\begin{equation}
    X_1=\frac{\kappa_2D_1-\kappa_1D_2}{D_1+D_2}, \;\;\; |\psi_1|^2=|\psi_2|^2=\frac{D_1+D_2}{\kappa_1+\kappa_2}-1, \;\;\; \Delta = \pm \sqrt{g^2-\left|\frac{\kappa_2D_1-\kappa_1D_2}{D_1+D_2}\right|^2}.
\end{equation}
For PT-broken solutions, \equref{eq:x1dependence} is more complicated, but still gives a direct formula for $D_2$ as a function of $X_1$. 

It is important to note that, for the previous solutions to be above threshold, we need $|\psi_1|>0$, or equivalently $D_1/(\kappa_1+X_1)>1$.

\bigskip
An exceptional point is obtained at the boundary between the two types of solutions. This occurs for $X_1=-X_2=\pm g$ and $\Delta = 0$. Plugging in \equref{eq:x1dependence} gives the condition for an exceptional point
\begin{equation}\label{eq:EP-condition}
    D_2 = \frac{\kappa_2\mp g}{\kappa_1\pm g}D_1.
\end{equation}
This requires $\kappa_1\geq g$ or $\kappa_2\geq g$. Each of the two corresponding exceptional points is above threshold (lasing) when $D_1 \geq \kappa_1\pm g$. One way to obtain the exceptional point is to first increase $D_1$ to a fixed value larger than $\kappa_1\pm g$, then increase $D_2$ to the required value for the EP.

\bigskip
We can also compute the output power of the laser as $P_{out} = \kappa_{1,r}|\psi_1|^2+\kappa_{2,r}|\psi_2|^2$, where $\kappa_{n,r}$ is the radiative loss rate. For simplicity, we consider the case of equal losses between the two resonators ($\kappa_1=\kappa_2$, $\kappa_{1,r}=\kappa_{2,r}$). For the PT-symmetric modes, we have 
\begin{equation}
    P_{out} = 2\kappa_r|\psi_1|^2 = \frac{\kappa_r}{\kappa}(D_1+D_2) -2\kappa_r.
\end{equation}
So the output power increases linearly with the pumps, with a rate $\partial P_{out}/\partial D_2=\kappa_r/\kappa$. The dependence of $P_{out}$ on the pumps is however more complicated for the PT-broken mode. We can still check that, at the EP, we have a discontinuity in the rate of $P_{out}$ by explicitly computing $\partial P_{out}/\partial D_2$. After some algebra, we find for the PT-broken mode at the EP
\begin{equation}\label{eq:ratepower}
    \frac{\partial P_{out}}{\partial D_2} = \kappa_r\frac{\kappa D_1-(\kappa\pm g)^2} {\kappa^2D_1-(\kappa^2-g^2)(\kappa\pm g)}.
\end{equation}

We then have a discontinuity in $\partial P_{out}/\partial D_2$ at the EP. The largest discontinuity is obtained for the smallest possible $D_1$ (i.e., $\kappa\pm g$). In the limit of large $D_1$, there is no discontinuity, as \equref{eq:ratepower} goes to $\kappa_r/\kappa$ for $D_1\rightarrow\infty$.
\bigskip

Our results here can also be used to interpret the experiments in \citeasnoun{kim2016direct} showing an EP laser in coupled photonic-crystal cavities. There, changing the position of the optical pump is equivalent to tuning $D_2/D_1$. But since the size of the pump-laser spot is large compared to the separation between the two cavities, $D_2/D_1$ could only be tuned slightly. The authors of \citeasnoun{kim2016direct} used a graphene sheet to partially cover one of the two cavities and increase its loss factor (equivalent to $\kappa_2$). In absence of the graphene sheet, the small loss gives a negative EP-ratio $D_2/D_1$ [\equref{eq:EP_condition}] so the laser stays in the PT-symmetric phase for any (positive) pump value [Fig.~4(d) of \citeasnoun{kim2016direct}]. With a large-area graphene sheet, $\kappa_2$ is large, giving a large EP-ratio $D_2/D_1$, so the laser remains in the PT-broken phase [Fig.~4(e) of \citeasnoun{kim2016direct}]. Finally, a small-area graphene sheet gives an intermediate value for the EP-ratio $D_2/D_1$ allowing for the observation of an EP at the transition between the two phases by tuning the pump position [Fig.~4(f) of \citeasnoun{kim2016direct}].

\section{Single-mode solutions}

\begin{figure}[!htp]
    \includegraphics[width=0.78\columnwidth, keepaspectratio]{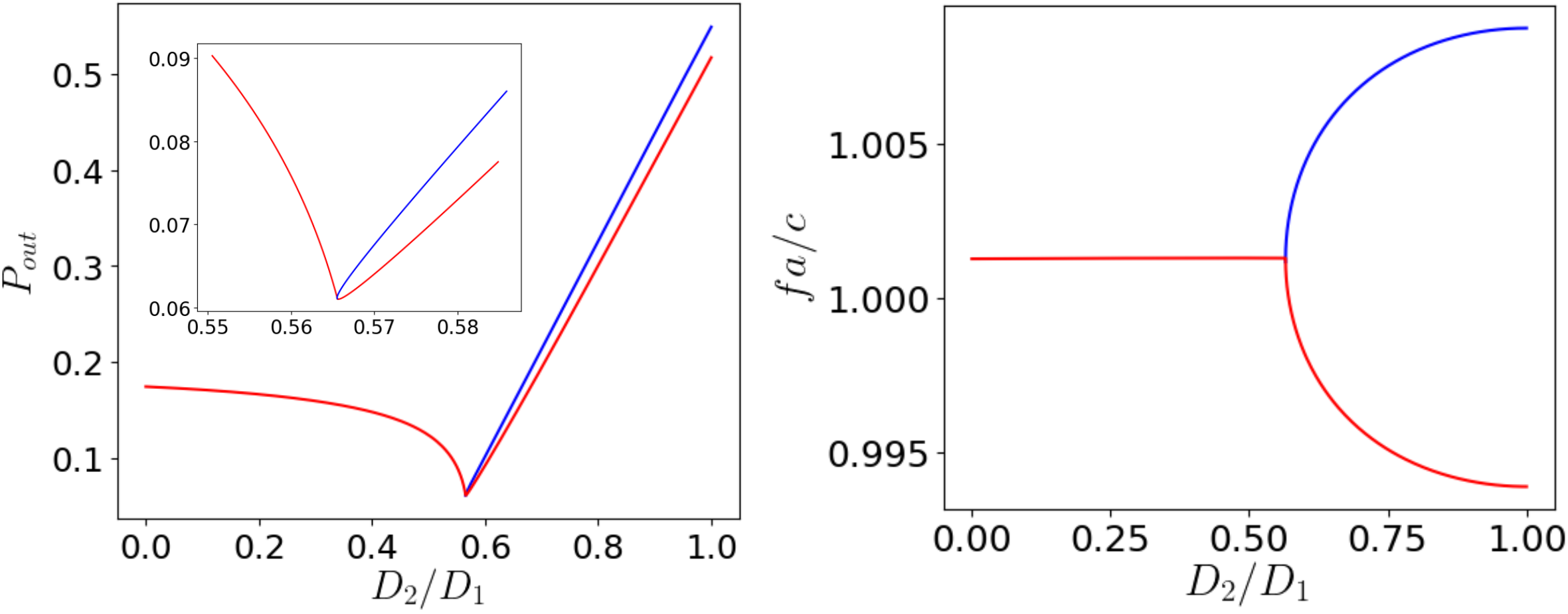}
    \caption{Laser output power (left) and frequency (right) for single-mode steady state solutions of the nonlinear Maxwell-Bloch equation, by keeping the first slab pumped at $D_1=0.85$. Compared to the main text, here we also show the second lasing mode in the PT-symmetric region (blue curve). The two curves are continuously joint through a fold point bifurcation. In practice, the choice of which mode actually lases (or possibly both) depends on some symmetry breaking factor and initial conditions.}
    \label{Fig_virtual}
\end{figure}

\section{Virtual nonlinear exceptional point}

\begin{figure}[!htp]
    \includegraphics[width=0.8\columnwidth, keepaspectratio]{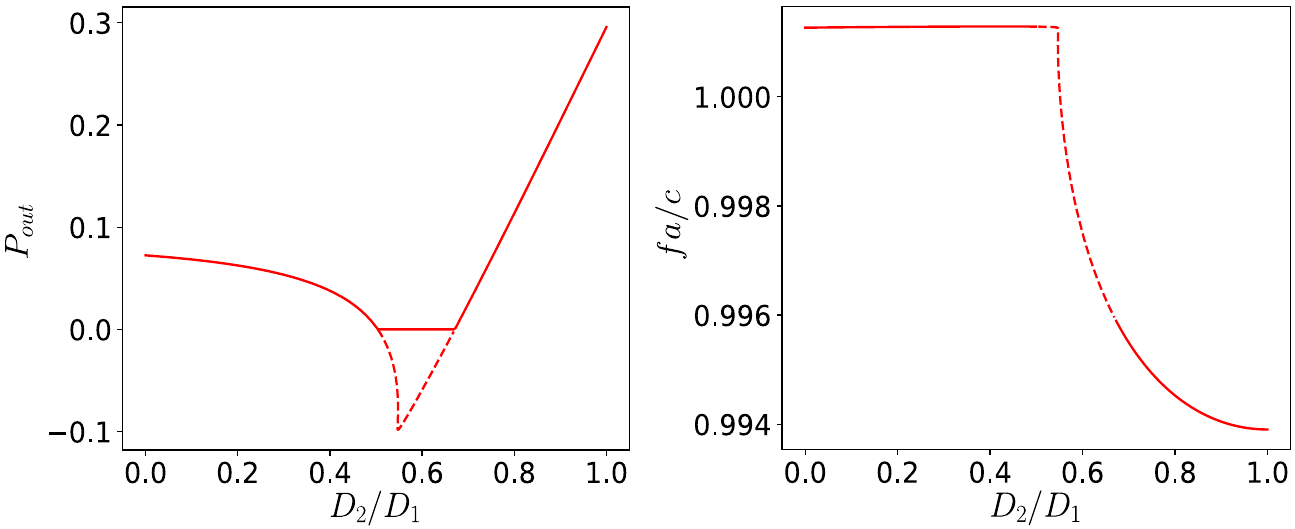}
    \caption{Laser output power (left) and frequency (right) for a single-mode steady state solution of the nonlinear Maxwell-Bloch equation, by keeping the first slab pumped at $D_1=0.85$. Dashed lines represent formal numerical solution with a gain saturation $D_\ell=D_0/\left(1+I_\ell|\Gamma_\ell E_\ell|^2\right)$ allowing $I_\ell$ to be negative. Physically, the laser shuts down when the power reaches zero (solid lines). At the lasing threshold the non-trivial negative intensity solution passes through the trivial non-lasing solution causing a bifurcation. This is true for any lasing mode and is not a special property of a mode at an EP~\citenum{cerjan2019multimode}.}
    \label{Fig_virtual}
\end{figure}

We consider the coupled-slabs cavity described in the main text. We solve the steady-state Maxwell-Bloch equation [Eq.~(7) of the main text] by pumping the first slab up to $D_1=0.85<\kappa+g\approx0.9$. To find non-zero $E_\ell$ solutions numerically, the nonlinear gain is written as $D_\ell=D_0/\left(1+I_\ell|\Gamma_\ell E_\ell|^2\right)$, where $I_\ell$ represents the laser intensity and the mode profile $E_\ell$ is normalized. While a physically meaningful solution requires $I_\ell$ to be positive, formal numerical solutions can still be obtained otherwise. This is shown as dashed lines in \figref{Fig_virtual}. Physically, the laser shuts down when $I_\ell$ reaches zero (solid lines). However, formal tracking of the laser field at ``negative power'' allows to find a \emph{virtual} nonlinear EP that explains the previously observed suppression and revival of lasing.   

\section{Linearized Maxwell-Bloch equations}

For a steady-state solution $(\mathbf{E}_\ell e^{-i\omega_\ell t}, \mathbf{P}_\ell e^{-i\omega_\ell t}, D_\ell)$ of Maxwell-Bloch equations, it has previously been shown that linearizing for additional perturbation $(\delta \mathbf{E} e^{-i\omega_\ell t}, \delta \mathbf{P} e^{-i\omega_\ell t}, \delta D)$ gives $\left(C\frac{d^2}{dt^2}+B\frac{d}{dt}+A \right)u(\vec{x},t) = 0$, where $u = (\text{Re}(\delta \mathbf{E}), \text{Im}(\delta \mathbf{E}), \text{Re}(\delta \mathbf{P}), \text{Im}(\delta \mathbf{P}), \delta D)$ and $A$, $B$ and $C$ are real operator matrices defined in previous literature~\cite{burkhardt2015steady, liu2017symmetry, benzaouia2020single}. Looking for solutions of the form $u=\text{Re}(U e^{\sigma t})$ leads to a quadratic eigenproblem 
 \begin{equation}\label{eq:instab-eigen}
 \left( A+B\sigma + C\sigma^2 \right)U = 0. 
 \end{equation} 
The sign of $\text{Re}(\sigma)$ determines the stability of the steady-state solution.

\begin{figure}[!htp]
    \includegraphics[width=\columnwidth, keepaspectratio]{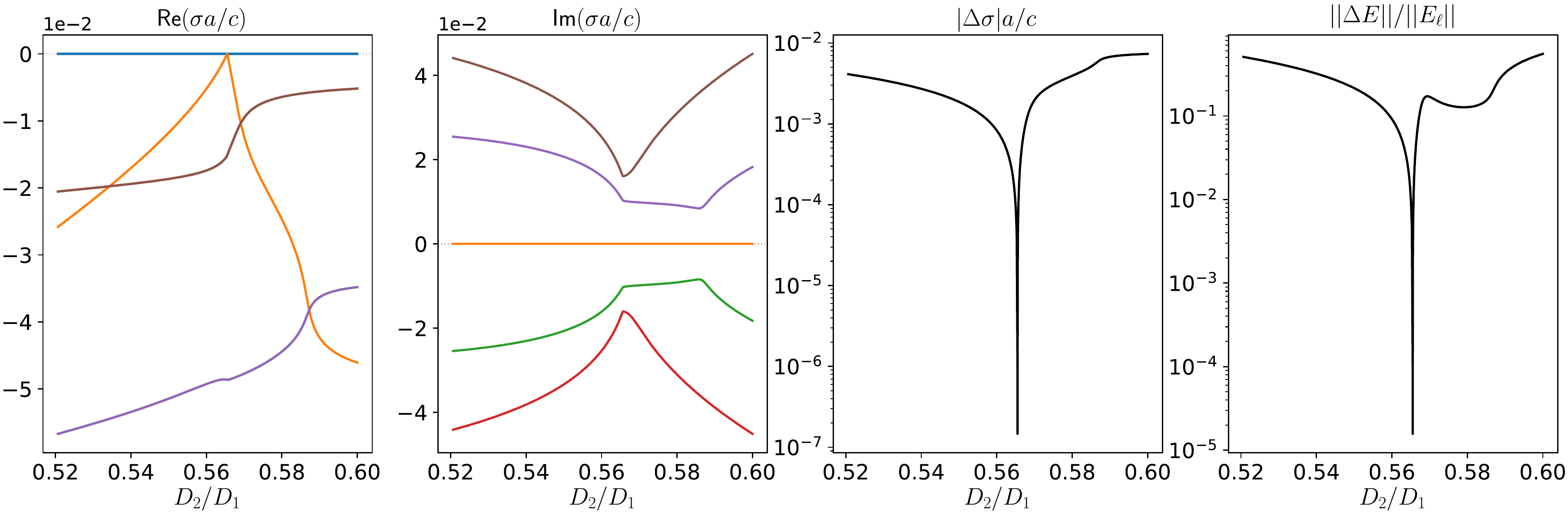}
    \caption{(a,b) Some eigenvalues of \equref{eq:instab-eigen}. (c,d) Norm of the difference between the two first eigenvalues/eigenvectors (the eigenvectors have been normalized to the same complex amplitude at the center of the structure).}
    \label{Fig_EPeigen}
\end{figure}

In \figref{Fig_EPeigen}(a,b), we show some solutions of \equref{eq:instab-eigen} for the structure considered in the main text and $\gamma_\parallel a/2\pi c = 10^{-2}$. The eigenvalues are either real or complex-conjugate (as expected since $A$, $B$ and $C$ are real operators). We clearly see that two eigenvalues coalesce at the EP. This is confirmed in \figref{Fig_EPeigen}(c,d) where we show the norm of the difference between the two eigenvalues and corresponding eigenvectors (where the eigenvectors have been normalized to the same complex amplitude at the center of the structure).  

\section{Scattering perturbation}

We assume a single-mode lasing solution to Maxwell-Bloch equations with field $E_\ell(x,t) = E_\ell(x)e^{-i\omega_\ell t}$. We scatter light off the laser cavity with an infinitesimal external source $\delta J_s(x,t)=\delta J_s(x)e^{-i\omega_st}$ at a frequency $\omega_s$. To solve this problem, we linearize Maxwell-Bloch equations around the lasing solution, which gives
\begin{equation} \label{eq:MB-linearization} \begin{split}
    \nabla^2 \delta E &=\delta \ddot P +\epsilon_c\delta\ddot E+\sigma_c\delta \dot E+\delta \dot J,\\ 
    \delta \dot P &= -(i\omega_a+\gamma_\perp)\delta P-i\gamma_\perp(D_\ell\;\delta E+\delta D\; E_\ell),\\
    \delta \dot D &= \gamma_\parallel\delta D+\frac{i\gamma_\parallel}{2}(\delta E\; P_\ell^*+E_\ell\;\delta P^*-\delta E^*\;P_\ell-E_\ell^*\;\delta P).
\end{split} \end{equation}
The presence of the nonlinear terms means that an idler field with frequency $\omega_i=2\omega_\ell-\omega_s$ will be generated. Using $\sigma=\omega_s-\omega_\ell$, we consider solutions of the form
\begin{equation}
    \delta E = \delta E_se^{-i\omega_s t}+\delta E_ie^{-i\omega_i t}, \;\;\delta P = \delta P_se^{-i\omega_s t}+\delta P_ie^{-i\omega_i t}, \;\; \delta D = \delta D_s e^{-i\sigma t}+\delta D_i e^{i\sigma t}.
\end{equation}
Note that since $\delta D$ is real, $\delta D_i=\delta D_s^*$. Plugging in \equref{eq:MB-linearization}, and equating each frequency component, for $m=\{s,i\}$ and $\Gamma_m = \Gamma(\omega_m)$, we have
\begin{equation} \begin{split}
    -i\omega_m\;\delta J_m&=\nabla^2 \delta E_m +\omega_m^2\delta P_m +\omega_m^2\epsilon_c\delta E_m+i\omega_m\sigma_c\delta E_m,\\
    \delta P_m &= \Gamma_m\left(D_\ell\;\delta E_m+\delta D_m\; E_\ell\right),\\
    \left(\gamma_\parallel-i\sigma\right) \delta D_s &= \frac{i\gamma_\parallel}{2}\left(\delta E_s\; P_\ell^*+E_\ell\;\delta P_i^*-\delta E_i^*\;P_\ell-E_\ell^*\;\delta P_s\right).
\end{split} \end{equation}
Now using the expression of $\delta P_s$ and the fact that $P_\ell=\Gamma_\ell D_\ell E_\ell$, we can compute $\delta D_s$ as
\begin{equation}
    -2\left(i+\sigma/\gamma_\parallel\right)\delta D_s = \left(\Gamma_i^*-\Gamma_s\right)|E_\ell|^2\delta D_s+(\Gamma_\ell^*-\Gamma_s)D_\ell E_\ell^*\; \delta E_s+(\Gamma_i^*-\Gamma_\ell)D_\ell E_\ell\;\delta E_i^*.
\end{equation}
We can finally use this to compute $\delta P_m$ as a function of the electric fields
\begin{equation}\begin{split}
    \delta P_s = \Gamma_s D_\ell \left[\left(1+\alpha |E_\ell|^2\right) \delta E_s + \beta E_\ell^2\; \delta E_i^*\right],& \; \; 
    \delta P_i = \Gamma_i D_\ell \left[\alpha^* E_\ell^2\; \delta E_s^* + \left(1+\beta^* |E_\ell|^2 \right)\delta E_i\right],\\
    \alpha = \frac{\Gamma_s-\Gamma_\ell^*}{2\left(i+\sigma/\gamma_\parallel\right)+\left(\Gamma_i^*-\Gamma_s\right)|E_\ell|^2},& \; \;
    \beta = \frac{\Gamma_\ell-\Gamma_i^*}{2\left(i+\sigma/\gamma_\parallel\right)+\left(\Gamma_i^*-\Gamma_s\right)|E_\ell|^2}.
\end{split} \end{equation}
We can finally plug in \equref{eq:MB-linearization} obtain coupled equations for $\delta E_s$ and $\delta E_i$
\begin{equation}\begin{split}
    -i\omega_s\;\delta J_s&=\nabla^2 \delta E_s +\omega_s^2\left[\epsilon_c+\frac{i\sigma_c}{\omega_s}+\Gamma_s D_\ell \left(1+\alpha |E_\ell|^2\right)\right]\delta E_s+\omega_s^2\; \Gamma_s D_\ell \; \beta E_\ell^2\; \delta E_i^*,\\
    0&=\nabla^2 \delta E_i +\omega_i^2\left[\epsilon_c+\frac{i\sigma_c}{\omega_i}+\Gamma_i D_\ell \left(1+\beta^* |E_\ell|^2\right)\right]\delta E_i+\omega_i^2\; \Gamma_i D_\ell\; \alpha^* E_\ell^2\; \delta E_s^*.
\end{split} \end{equation}

\end{document}